\documentclass[preprint,12pt]{article}
\pdfoutput=1 

\usepackage{xcolor}
\usepackage{jinstpub}
\usepackage[numbers]{natbib}
\usepackage{graphicx}

\usepackage{url}

\usepackage{lineno}
\usepackage[version=4]{mhchem}
\usepackage{siunitx}

\newcommand{\geant}{\texttt{GEANT4}}
\newcommand{\ROOT}{\texttt{ROOT}}
\newcommand{\iso}[2]{\ce{^{#2}#1}}
\newcommand{\xe}[1]{\iso{Xe}{#1}}

\title{Simulated Thick, Fully-Depleted CCD Exposures Analyzed with Deep Learning Techniques }

\author[a]{ C. Britt,}
\author[a]{ E. Church,}
\author[a]{T. Hossbach,}
\author[a]{B. Loer,}
\author[a]{R. Saldanha,}
\author[a]{ N. Sinha,}
\author[b]{K. Woodruff}
\affiliation[a]{Pacific Northwest National Laboratory, Richland, WA 99354}
\affiliation[b]{University of Texas - Arlington, Arlington, TX }
\emailAdd{carl.britt@pnnl.gov, eric.church@pnnl.gov, todd.hossbach@pnnl.gov,  ben.loer@pnnl.gov, richard.saldanha@pnnl.gov,  nishitasinha@college.harvard.edu, katherine.woodruff.2@gmail.com}

\abstract{
Thick, Charge Coupled Devices (CCDs) have recently been explored for applied physics, such as nuclear explosion monitoring, and dark matter detection purposes. When run in fully-depleted mode, these devices are sensitive detectors for  energy depositions by a variety of primary particles. In this study we are interested in applying the Deep Learning (DL) technique known as panoptic segmentation to simulated CCD images to identify, attribute and measure energy depositions from radioisotopes of interest. We simulate CCD exposures of a chosen radioxenon isotope, $^{135}$Xe, and overlay a simulated cosmic muon background appropriate for a surface-lab. We show that with this DL technique we can reproduce the beta spectrum to good accuracy, while suffering expected confusion with same-topology gammas and conversion electrons and identifying cosmic muons less than optimally.
}

\begin{document}
\maketitle


\tableofcontents

\section{Introduction: CCDs for radiation detection}
    Thick Charge Coupled Devices (CCDs) have recently shown a potential usefulness for dark matter detection~\cite{aguilar-arevalo2015} and for applied physics where it is desired to detect daughters of radioisotopes. Energy resolution is very high and particle energy deposition topologies are distinct and identifiable.
    
    Particle physics experiments have used various types of semantic segmentation to classify  pixels~\cite{ubss,ubpid} and even perform object identification~\cite{uboi}, but not in CCDs. We show here the first application, to our knowledge, of Deep Neural Networks (DNNs) for particle identification (PID) in simulated CCDs exposed to radioactive gas samples. In the conventional application, simple clustering algorithms operate by starting at a seed point and clustering all contiguous energy depositions. Any PID is done by examining energy deposition intensity. In this paper we use the learned DNN PID to show that the component energy spectra give a good approximation to the full ones, suggesting that future PID applications -- such as actually inferring an unknown gas isotope -- in data may be possible. For now this exercise is beyond the scope of this study and left for future demonstration. 
    
    Our goal here is to use a technique called panoptic segmentation to identify potentially multiple instances of various classes of objects in an image and then to label the relevant pixels as belonging to those classes. The simpler semantic segmentation task is more commonly exercised. Semantic segmentation does the  work of identifying pixels in an image, like those attributable to cats, parakeets or humans but without awareness of instances of each cat, parakeet or human in the image. Whereas, if we want to find all the unique cats, dogs, humans, parakeets in an image, we want to apply panoptic segmentation. To do this, we apply an object detection network in this paper in order to find the betas from $^{135}$Xe decay and the subsequent conversion electrons (CEs) and de-excitation gammas, and separately the background muons from cosmic rays.
    
    We will also create and study a diagnostic sample of electrons to explore some observations we make in the \xe{135} sample, which is useful in nuclear explosion monitoring. This sample will be of a relatively narrow energy range and is designed to help us investigate the classifiability of gammas and electrons that are relevant to our radioxenon studies.
    
\begin{figure}
        \centering
        \includegraphics{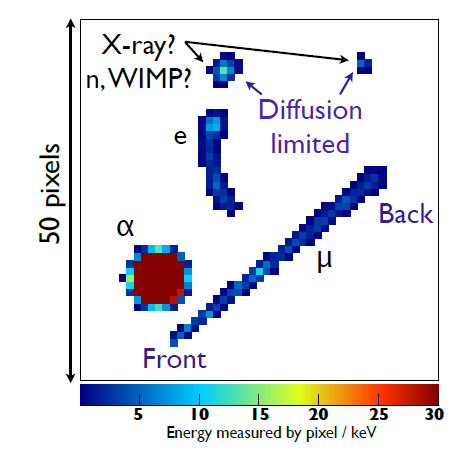}
        \caption{Example of typical track topologies for different particle interactions as measured by thick CCDs for the Dark Matter In CCDs (DAMIC) experiment~\cite{aguilar-arevalo2015}.}
        \label{fig:ccdtracks}
    \end{figure}

    As Figure~\ref{fig:ccdtracks} demonstrates, there are four qualitatively distinct radiation track types produced in the CCD image by radioactive decay particles:
    \begin{enumerate}
        \item Low energy photon processes (Compton scattering or photoelectric effect), where the track of the resulting electron is sub-pixel length, produce diffusion-limited points
        \item Higher energy electrons (from Compton scattering, beta decay, or conversion electrons) produce extended meandering tracks
        \item Cosmic ray muons produce extended straight line tracks
        \item Alpha particles produce large bright circular features. Protons  produced by cosmic rays produce similar, smaller features. Alphas are the least relevant decay signature for the studies in this paper.
    \end{enumerate}
    \xe{135} is a useful test case because it produces both extended and point-like electron tracks. 
    
\section {Simulated energy deposition of radioisotopes}

    The first step in generating artificial images with known truth information is to model the transport of particles to generate energy deposition tracks in the silicon substrate. A model of a thick 6k$\times$1k pixel CCD was constructed using the \geant\ Monte Carlo toolkit~\cite{agostinelli2003}, version 10.06.p02. The CCD is modeled as a 95$\times$16~mm silicon wafer, 675~$\mu$m thick. The front side has a 2~$\mu$m thick \ce{SiO2} dead layer.  The back side has a 50~nm dead layer and a 100~nm aluminum conductive layer.  The active area is 92.16$\times$15.36~mm corresponding to 6144$\times$1024 15~$\mu$m pixels. 
    
    For particle tracking, the base physics list \texttt{FTFP\_BERT} is used with the high precision options \texttt{EmOption4} for low energy charged particles. Within the silicon the \texttt{MicroElec} precision options are also enabled~\cite{valentin2012}.  During each step of particle tracking, the total energy deposited is recorded along with three dimensional position coordinates. The maximum step length is limited to 5~$\mu$m (i.e. one-third of a pixel side) so that each step may be treated as a point in the image generation process described in the next section. 
    
    We simulated two types of primary particles: cosmic ray muons and radioactive xenon isotopes (radioxenon).  Primary muons were generated from a square plane 80~m on a side, 20~cm above the CCDs. The initial energy and angular distributions were generated using the \texttt{CRY} cosmic ray shower generator library~\cite{hagmann2007}. The CCD may be rotated so that the face is parallel or perpendicular to the geometry's vertical $z$ axis, which affects the average muon track length. For the images used in this work, the CCD was oriented with the face parallel  to the $z$ axis (the majority muon direction), resulting in longer average muon tracks. 
    
    For radioxenon sources, we simulate a hypothetical gas cell in front of the CCDs. The cell is 1~cm thick, 1~mbar of \ce{CO2}. Radioactive isotopes of xenon (\xe{131m}, \xe{133}, \xe{133m}, \xe{135}) are distributed uniformly throughout the gas volume and allowed to decay via the built-in radioactive decay of \geant{}. The gas cell can be placed on the front or back side of the CCDs. After the primary isotope decays, any prompt deexcitation of the daughter also proceeds, but decays into long-lived radioactive daughters (e.g. \iso{Cs}{135} produced by \xe{135}) are suppressed.  The direct parent of each radiated particle is recorded with each energy deposition to provide additional MCTruth information (e.g. beta decays emitted by xenon isotopes vs x-rays and conversion electrons emitted by the daughter nucleus). The images of radioxenon sources also include muon tracks as background. Typical images were generated by randomly selecting 120 \xe{135} decay events and 30 muon events, corresponding to an exposure time of approximately 6 minutes (determined by the typical rate of muons at the surface) and a radioxenon activity level of approximately 1 Bq. This rate was chosen to minimize the problem of overlapping energy depositions, which demands more of the object identification, while still representing a realistic activity of a typical lab gas sample.
    
    We also create and study a diagnostic sample of electrons with kinetic energy between 100 and 200 keV. These are launched isotropically from the front, the back, and the bulk of the CCD in equal proportion. For this sample we want to explore both the classification accuracy of these samples with relatively controlled start-point and diffusion distances -- the bulk electrons mimicking gammas that convert into the CCD -- and absent the context of the full radioxenon decay chain.
    
\section {Charge drift and diffusion model and MCTruth Information}
   
The next step in the generation of artificial CCD images is to model the CCD response to the simulated \geant\ energy deposits. 

\subsection{Signal Generation}
For each simulated energy deposit, the particle type, energy, and the 3-D location are obtained from the \geant~simulation. The location is calculated by averaging the start and end locations of the \geant~step. We then perform the following simulation steps.
\subsubsection {Charge Generation}
The number of electron-hole pairs corresponding to the energy deposit is randomly drawn from a Gaussian distribution. The mean of the distribution is set to be the energy deposited divided by the mean energy to make an electron-hole pair (3.6 eV), while the standard deviation is set to correspond to a Fano Factor of 0.16 \cite{janesick2001scientific}.
\subsubsection{Diffusion}
The diffusion of charge along the transverse direction (perpendicular to the drift) is determined by the depth of the interaction and the applied bias voltage. We use an empirical model for the width of the diffusion $\sigma$ as a function of depth $z$ given by
\begin{align}
\sigma^2(z) = -A\ln(1-bz)
\label{eq:diffusion_width}
\end{align}
where $A =$ \SI{254}{\micro\meter\squared} and $b = $ \SI{8.18E-4}{\per\micro\meter} \cite{aguilar-arevalo2015}.

To simulate the pixelation of the CCD readout, the top surface ($z=0$) of the silicon substrate is divided into \SI{15}{\micro\meter} $\times$ \SI{15}{\micro\meter} pixels corresponding to a total of 6k $\times$ 1k = \num{6.29E6} pixels in the $x-y$ plane. The diffusion is applied using an image kernel or convolution matrix with a Gaussian kernel. Since the width of the Gaussian kernel depends on the depth ($z$) of the energy deposit and the location of the energy deposit in the $x-y$ plane is not necessarily aligned with the pixels, the kernel must be recalculated for each energy deposit. We have chosen a $7\times7$ pixel kernel with the center at the pixel closest to the location of the energy deposit. The kernel values at each pixel are then calculated by integrating a normalized 2-D Gaussian, whose width is determined by Eq.~\ref{eq:diffusion_width}, over the pixel area. For an energy deposit at $(x_c, y_c, z_c)$ and pixel at $(x_p, y_p)$, the kernel is calculated as
\begin{align}
    k(x_p, y_p) = \frac{1}{4}
    \left(\text{Erf}\left(\frac{x_c - (x_p-\delta x)}{\sqrt{2}\sigma(z_c)}\right) + \text{Erf}\left(\frac{(x_p+\delta x) -x_c}{\sqrt{2}\sigma(z_c)}\right)\right) \nonumber\\
    \left(\text{Erf}\left(\frac{y_c - (y_p-\delta y)}{\sqrt{2}\sigma(z_c)}\right) + \text{Erf}\left(\frac{(y_p+\delta y) -y_c}{\sqrt{2}\sigma(z_c)}\right)\right)
\end{align}
 where $\delta x$ and $\delta y$ are the half-widths of the pixels in the $x$ and $y$ dimensions respectively.
 
 The value of the kernel is then added to the corresponding pixel in the signal image for the specific particle type. The images  for each of individual particle types are saved in the format of a \ROOT~\cite{brun1997root} 2-D histogram (\texttt{TH2F}) for ease of manipulation and display. We note that kernel pixels that fall outside the bounds of the silicon substrate are set to zero, with that fraction of the charge being lost.

\subsection {Noise Generation}
Independently of the signal generation, a separate noise image (2D histogram) is generated for each CCD exposure being simulated. There are two sources of noise considered.
\subsection{Dark Current}
We have assumed a mean dark current rate of \SI{1E6}{e^-\per\hour\per pixel} with a default exposure time set to 6 minutes. The specific value of the dark current in each pixel is randomly and independently drawn from a Poisson distribution with the above dark current rate. 
\subsection{Readout Noise}
The noise generated by the readout of each pixel is simulated as a random variable drawn from a Gaussian distribution whose mean is \SI{0}{e^-} and whose width is \SI{1.6}{e^-}. In the current version of the simulation the readout noise from each pixel is considered independent and uncorrelated to other pixels.

\begin{figure}
    \centering
    \includegraphics[width=\textwidth]{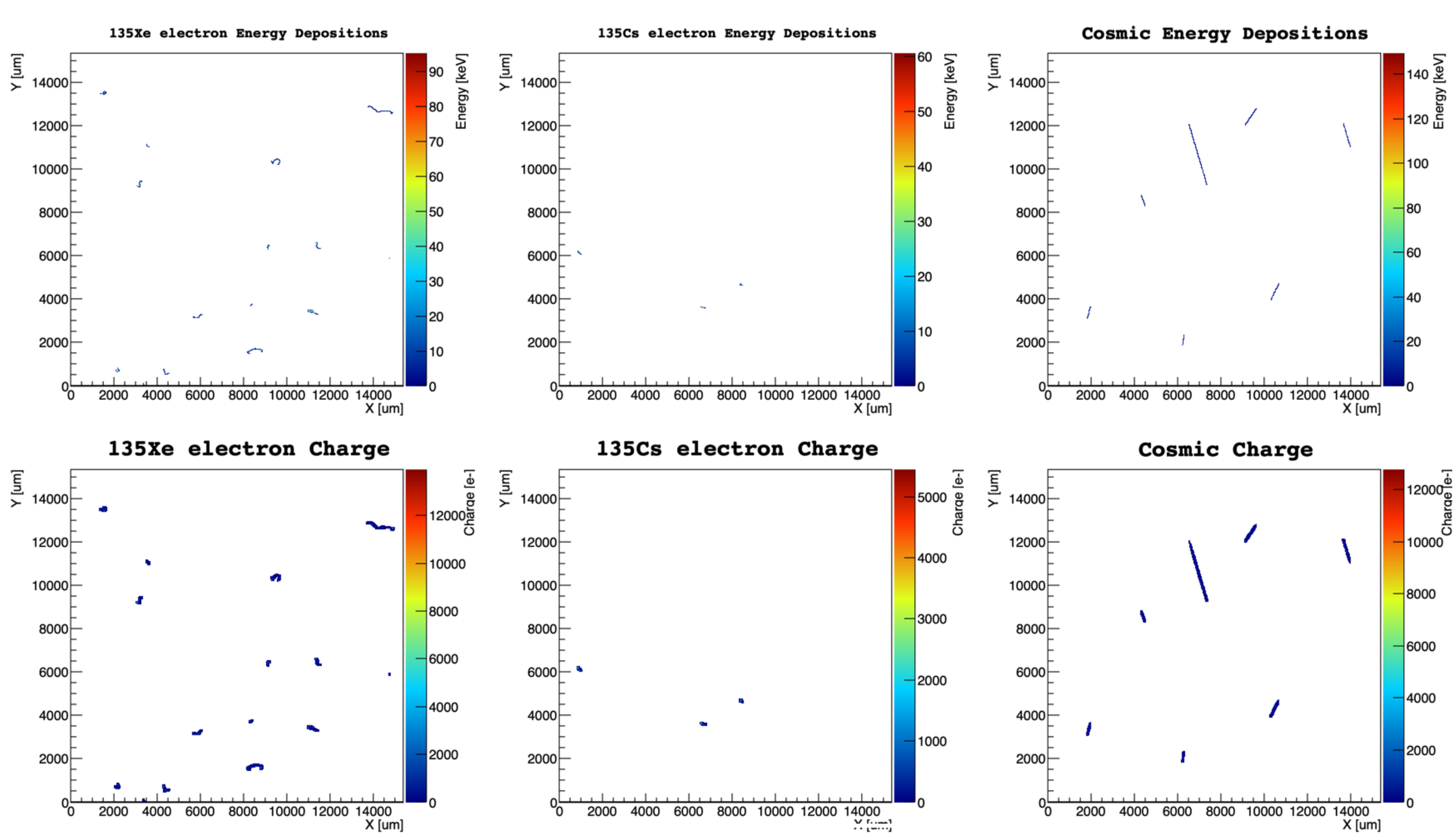}
    \includegraphics[width=\textwidth]{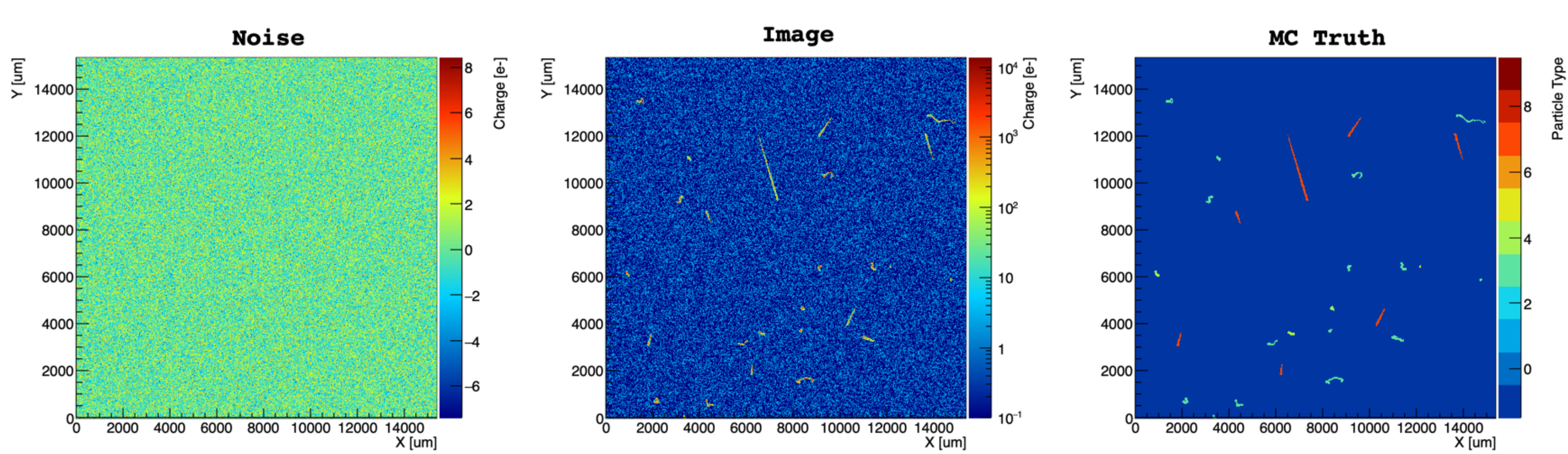}
    \caption{Example set of simulated CCD images for a single exposure combining \xe{135} and cosmic ray muons. The top row shows the energy deposits for three particle types from the \geant~simulation. The second row shows the corresponding simulated charge produced in the CCD, after including the effect of diffusion. The bottom left figure shows the simulated noise image. The bottom center image shows the final simulated image that combines the charge from all the different particles as well as the noise. The bottom right image shows the MC Truth information for each pixel in the final image. The legend for particle type is shown in Table~\ref{tab:mctruth_legend}.}
    \label{fig:sim_images}
\end{figure}
\subsection{Image Generation and Monte Carlo Truth}
The final image is created as a 2D histogram (\texttt{TH2F}) with each pixel (bin) value set as the sum of the corresponding pixel values from each of the particle type images and the noise image.

For each image we create a corresponding Monte Carlo Truth image in which each pixel is labeled by the largest contributor of charge to the pixel. As shown in the example for \xe{135} in  Table~\ref{tab:mctruth_legend}, the possible contributors are labeled by parent isotope and particle type, or as noise if it is the dominant contributor.

\begin{table}[h]
\centering
\begin{tabular}{ l c }
Charge source  & MC Truth Label \\
\hline
Noise & 0\\
\xe{135} $\beta$ & 4 \\ 
$^{135}$Cs Conversion electron & 5 \\
$^{135}$Cs De-excitation photon & 6 \\
Cosmic muon-related & 8
\end{tabular}
\caption{Possible sources of charge in each pixel for \xe{135} images and the corresponding MCTruth integer labels applied to each pixel depending on the largest contributor of charge.}
\label{tab:mctruth_legend}
\end{table}

Figure~\ref{fig:sim_images} shows an example set of intermediate simulated images and the corresponding final CCD image an MC Truth for a single exposure of \xe{135} and cosmic ray muons. 
    

\section {Deep Neural Networks}

We have adapted the Detectron2~\cite{detectron2} network to our application. Detectron2 is typically used to identify objects, e.g., people, bicycles, cars and traffic signs in images, place bounding boxes around them and properly label all the pixels in the boxes. For this application, we have supplied the ground truth bounding box, mask, and energy deposition information for each particle interaction captured in the 2D histogram as the data set. 

\subsection{Our Particular Application }
We  discovered that our relatively small and numerous particle depositions require adjustments to the default Detectron2 configuration and training protocols. After such adjustments, training becomes apparent and performance in test samples remains robust. We measure our performance with the metric Panoptic Quality~\cite{pq} (PQ), which can be  interpreted as a product of purity and efficiency of identification, or recall and precision in the computer vision parlance. 
    
Detectron2 consists, in rough terms, of a network that simultaneously, and separately, finds "activation filters" and proposed box regions of activity. These proposed regions are then matched to ground truth bounding boxes where the pixel Intersection-over-Union (IoU) is high. The loss function to be minimized includes many terms: one to find the correct proposed regions in their right place, one for putting the correct pixel mask within the boxes, and another for the proper classification of pixels in the box. In this way actual desired foreground boxes are eventually separated from poorly guessed-at  background boxes, and those foreground boxes contain the mostly-correct pixels for the correct class. For  more details, the reader is directed to References~\cite{ren2015faster, lin2017feature, he2017mask} which describe individual components of Detectron2.
    
We have many parameters in Detectron2 at our disposal, of which we have availed of only some fraction. We acknowledge that results here are not yet optimal, and can be expected to improve with further such work. For an example of possibilities explored here, during training we may demand that only some of the activation regions contribute. Empirically here, we find best results from allowing all four possible activation regions. We also find while training that the "Pooler\_Resolution" setting of 1 for both the ROI and Features networks gives superior training, distinct from the default of higher values which is perhaps more suitable to finding macroscopic regions of interest (backpacks, signs, cars, etc.) in a picture, than the often tiny charged particle features we'd like to identify here. Next, we may ask during test sample running that only boxes where the predicted score exceeds some threshold be counted as true predictions. And subsequent to training and testing, during analysis we may choose our own IoU threshold of ground-truth/best prediction overlap to determine whether a prediction is correct or not. In the next section we make some of the above choices and show resulting performance trade-offs.
    
\subsection{Our Procedure}
Results are obtained on the Summit cluster at the Oakridge Leadership Class Facility, recognized at the time of performing this work as the fastest supercomputer in the world. We typically train for over 120 epochs over 12 wall-clock hours on 24 GPUs at a time.
    
With the simulated activity specified, we find that we must run with 25\% of our approximately 10cm x 2.5 cm CCD image at a time, so as to fit onto our GPUs. These quarter images contain approximately 1500x1000 pixels, and so remain large with respect to the typical photograph analyzed by panoptic segmentation networks.

Starting from our 300 initial simulated  exposures, we thus have effectively 1200 images, of which 80\% are used to train the network. The other 20\% comprise the test sample. Given the rather local nature of the ionizing particles studied here we see that quartering our images poses very little compromise to particle ID or spectrometry. There are typically 60$\pm 15$ energy depositions per 1500x1000 pixels with the \xe{135} we have imposed. 

We find that a batch size of only one or two such images  per GPU is permitted by memory constraints on the Summit NVIDIA V100s. However, with N GPUs and parallelization afforded by our software packages -- larcv~\cite{larcv} for data manipulation, and horovod~\cite{horovod} as the means to launch jobs to GPUs and aggregate the output -- we have a much larger effective batch size. Typically we employ as large as N=96 GPUs for training runs and N=24 for final test runs. Another big advantage  of our large effective batch size, beyond the stable learning achievable with such batches, is that we can complete one full epoch on Summit in typically 10 minutes or so. With user Summit jobs limited to two wall-clock hours, therefore, we see that development can proceed in reasonable time with such a Leadership Class Facility at our disposal. We generally, write out the model at the end of each job, and then run the next two-hour job, starting at the previous model. This submission process is automated. Ten or twenty such jobs yield the results shown here.

\section{Data preparation}

The network requires considerable care and labeling of the input in order to train. Ground truth information must be created and given to Detectron2 in a particular format.

\begin{figure}
        \centering
        \includegraphics{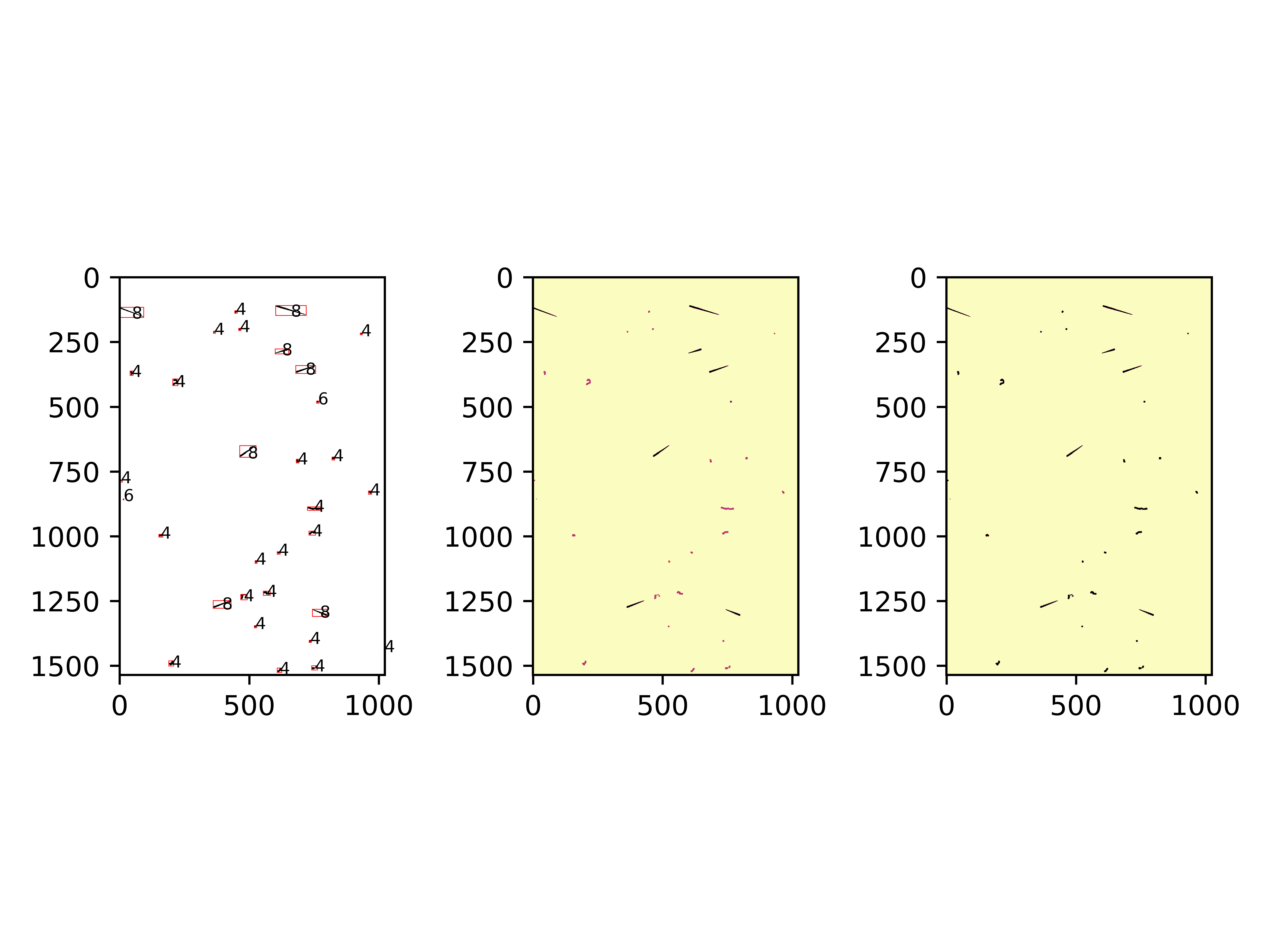}
        \caption{Example raw image (right), color-coded ground truth (middle), and bounded, truth-labeled (left). Note from left image that the energy depositions are dominantly 4s (\xe{135} betas) and 8s (cosmics), though a few 5s and 6s show that conversion electrons and X-rays are also present. Table~\ref{tab:mctruth_legend} summarizes these deposition truth labels.}
        \label{fig:labeling}
    \end{figure}

Ground truth bounding boxes are drawn bounding every unique particle trajectory, pixels in that bounding box are labeled as 1 or 0 according to whether they're above-noise-threshold or are merely background in the box, and the box is given an overall label. That label in our figure is an integer that identifies one of the four features of interest in our study. Figure~\ref{fig:labeling} shows three views of a particular event. The left pane of that Figure encapsulates the input that goes to the network. 

\section {Results}
Here we discuss results of the network training over the train sample and inference over the test sample.

\subsection {Training}
    
\begin{figure}
        \centering
        \includegraphics{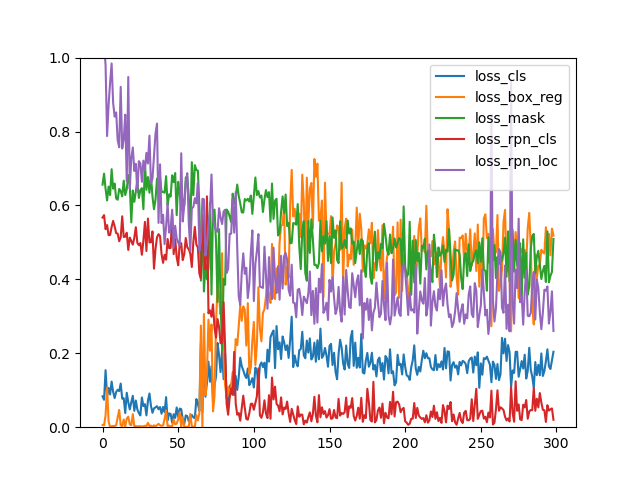}
        \caption{Example of loss function(s) versus epoch. The components labeled with "RPN" indicate the location and class correctness of the early-on regional proposal network, whereas the other three are for the final box, class, and mask (pixel-level) losses. Note the  trade-offs among the components (in which frequently the early-training bounding boxes are overly-large and sparse and by-eye clearly unuseful) and the relative stability in the final 100-ish epochs.}
        \label{fig:loss}
    \end{figure}

Figure~\ref{fig:loss} shows the loss trajectory of our network on the training sample. We use Detectron2's loss functions as-built; losses are calculated for five components and then summed. We inspect the components through the training epochs.

\subsection{Predicted Labels}

\begin{figure}
        \centering
        \includegraphics{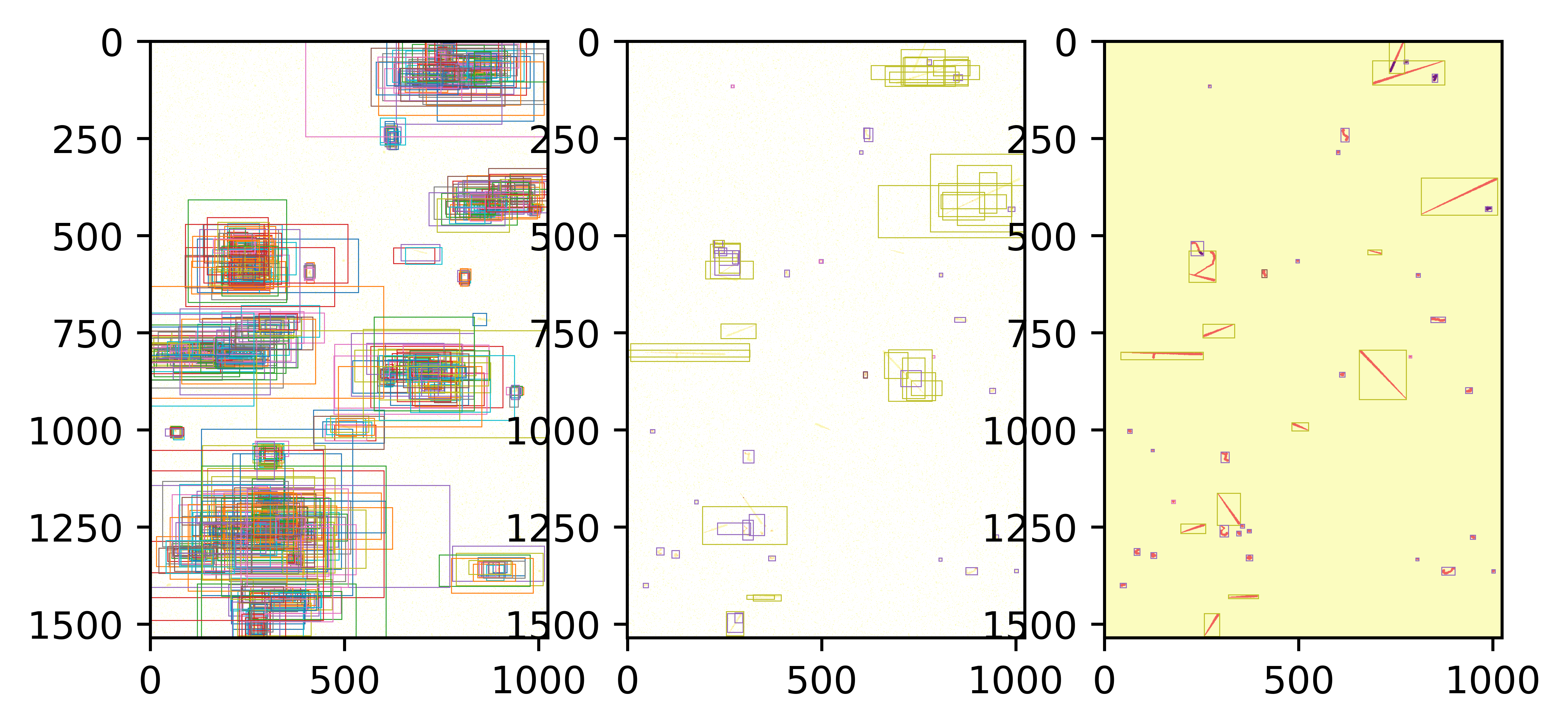}
        \caption{Shown are truth-labeled image in right hand pane, early-in-network Region Proposed bounding boxes in left hand pane, and final results in central pane. Minimum score to appear in middle pane is a relatively low 0.05. Purple boxes show betas, yellow show cosmics. Compared to right pane, the performance by-eye is not unreasonable, though it's evident the cosmics are redundantly boxed.}
        \label{fig:preds}
    \end{figure}

It helps to see the predictions of the network on an individual image after training is complete. Figure~\ref{fig:preds} shows such an image. The inference here only shows bounding boxes for which the score exceeds the default score of 0.05. One notes from a by-eye  comparison of the middle pane to the left pane a fairly complete set of boxes where they belong and roughly properly labeled, though there are many "redundant" boxes around cosmic activity.

\begin{figure}
        \centering
        \includegraphics{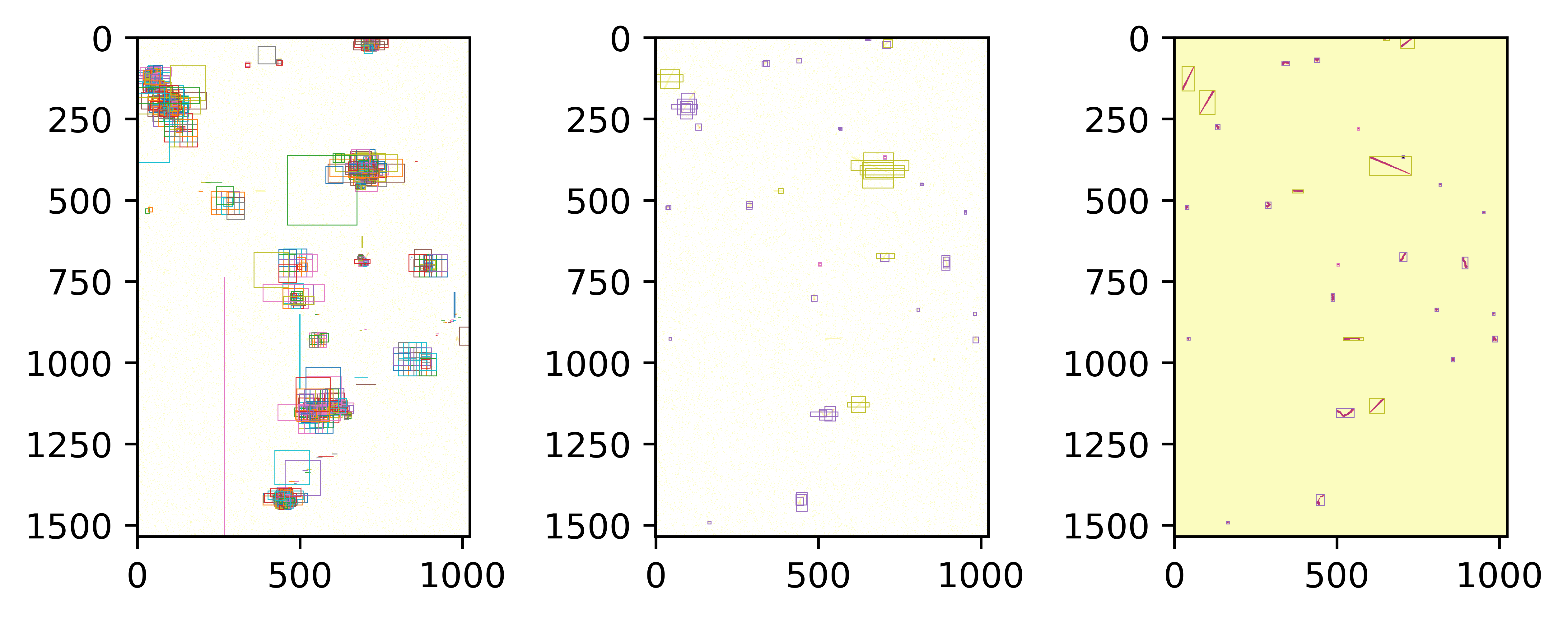}
        \caption{Shown are truth-labeled image in right hand pane, early-in-network Region Proposed bounding boxes in left hand pane, and final results in central pane. Minimum score to appear in middle pane is now 0.20. Purple boxes still show betas, yellow show cosmics. We see more activity here that goes unidentified.}
        \label{fig:preds20}
    \end{figure}

We show another (different) image in Figure~\ref{fig:preds20} for which the score for any identification must exceed the minimum of 0.20. One sees clear activity which is not identified (false negatives) and less, though not entirely eliminated, redundant bounding boxes. 

We quantify completeness and correctness of network predictions in the following subsections.

\subsection{Energy Spectra}

\begin{figure}
        \centering
        \includegraphics[width=4.0in]{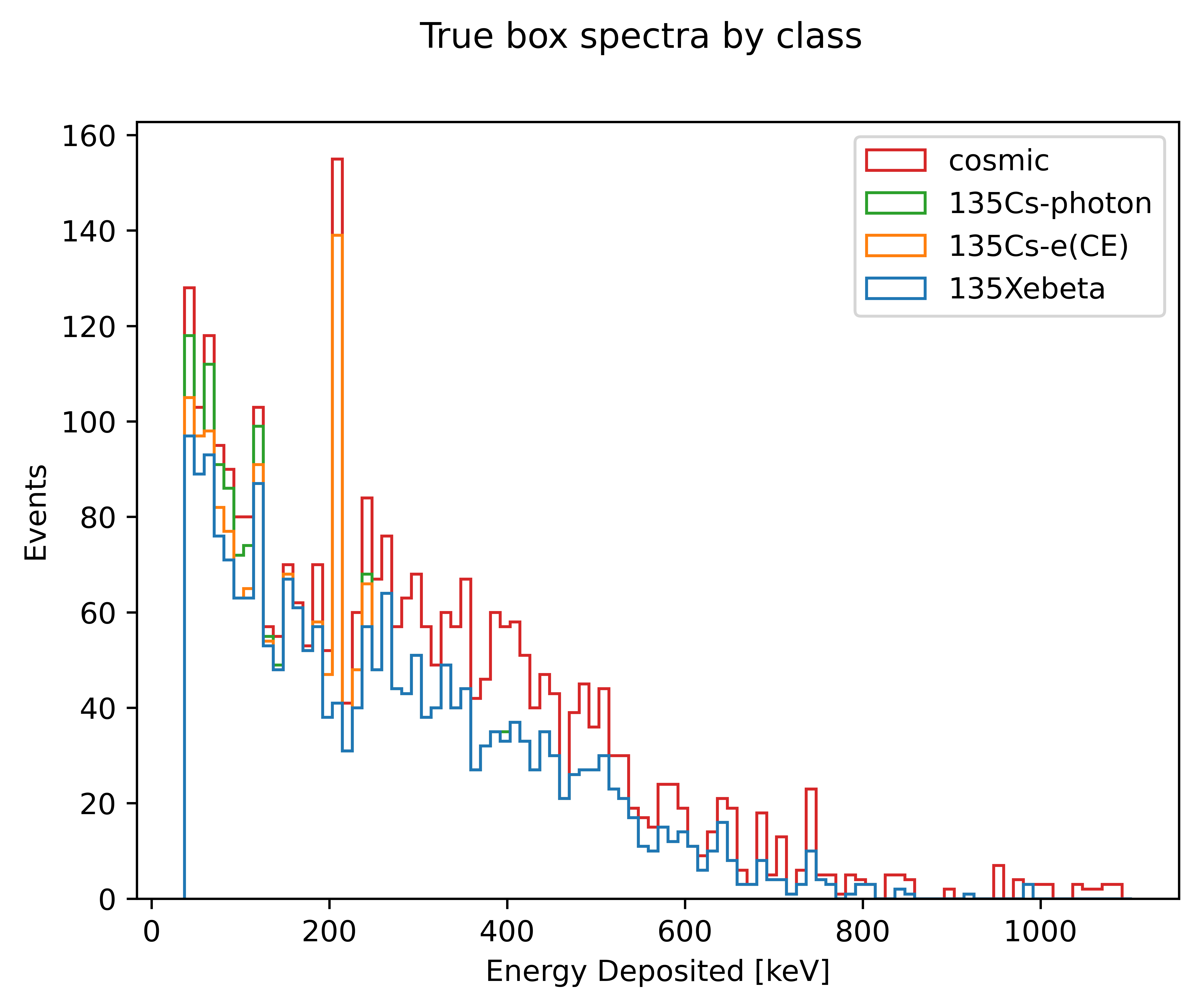}
        \includegraphics[width=4.0in]{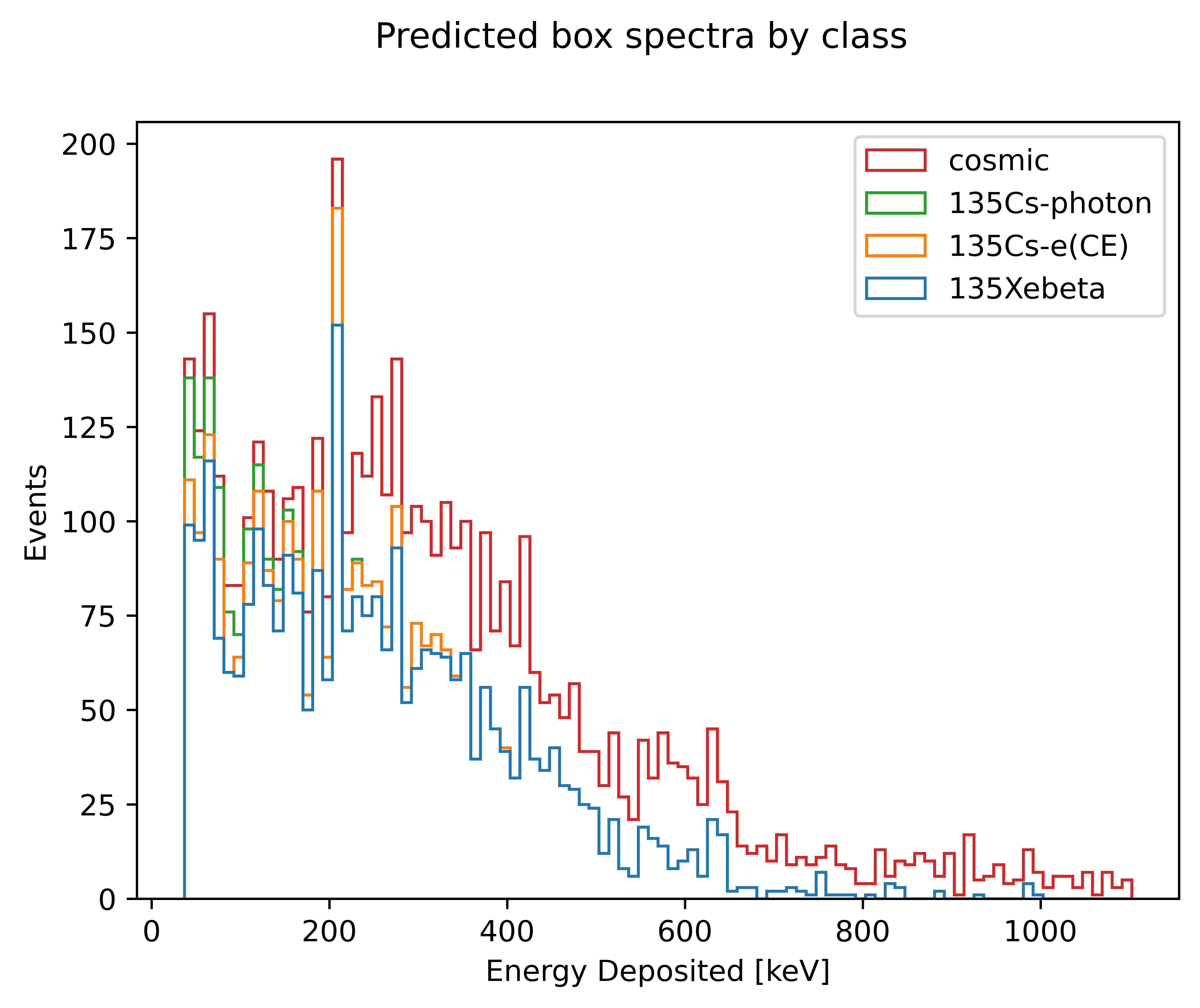}
 
        \caption{Shown are stacked histograms of calculated energy distributions for the ground-truth images on top, and the network predictions on bottom. Each image allows the four contributions shown. Minimum score to appear in prediction distribution is 0.05.}
        \label{fig:spectra}
    \end{figure}

We show in Figure~\ref{fig:spectra} the energy spectra of true and predicted images (from the test sample) again for  minimum predicted scores of 0.05. This is our main result. 

We note the evident beta activity, properly identified, as the dominant component of the histogram. On the downside, our most prominent conversion electron (CE) is properly identified some of the time and guessed at to be a beta others. Similarly, the very low energy X-ray is identified correctly about half the time. We remind at this point that the Compton electrons from the parent X-ray with a mean free path at 200 keV of about 5cm are as likely to occur anywhere below the face of the CCD as any other, so we may hope that their diffusion pattern might look slightly different than the low energy betas which start right at the gas interface with the CCD. In fact, we are heartened by the identification of some of this low energy activity, though from later metrics it will become clear our performance for these gammas isn't terribly impressive. As to the cosmic muons, they pile up about where they're expected but also erroneously down at lower energies. This is a result of the bounding boxes around our cosmics being insufficiently large coupled with the fact that we are predicting extraneous cosmics boxes.

We remark here that to produce our stacked energy histogram, we choose to sum all activity in a predicted box, rather than just summing the labeled pixels inside that box. It turns out the labeled pixel sum gives a  systematically too low spectrum with respect to that from the full box sum. At our chosen radioactivity we have almost no problem with overlapping boxes, and so this choice is non-worrying to impose.

\begin{figure}
        \centering
        \includegraphics[width=\textwidth]{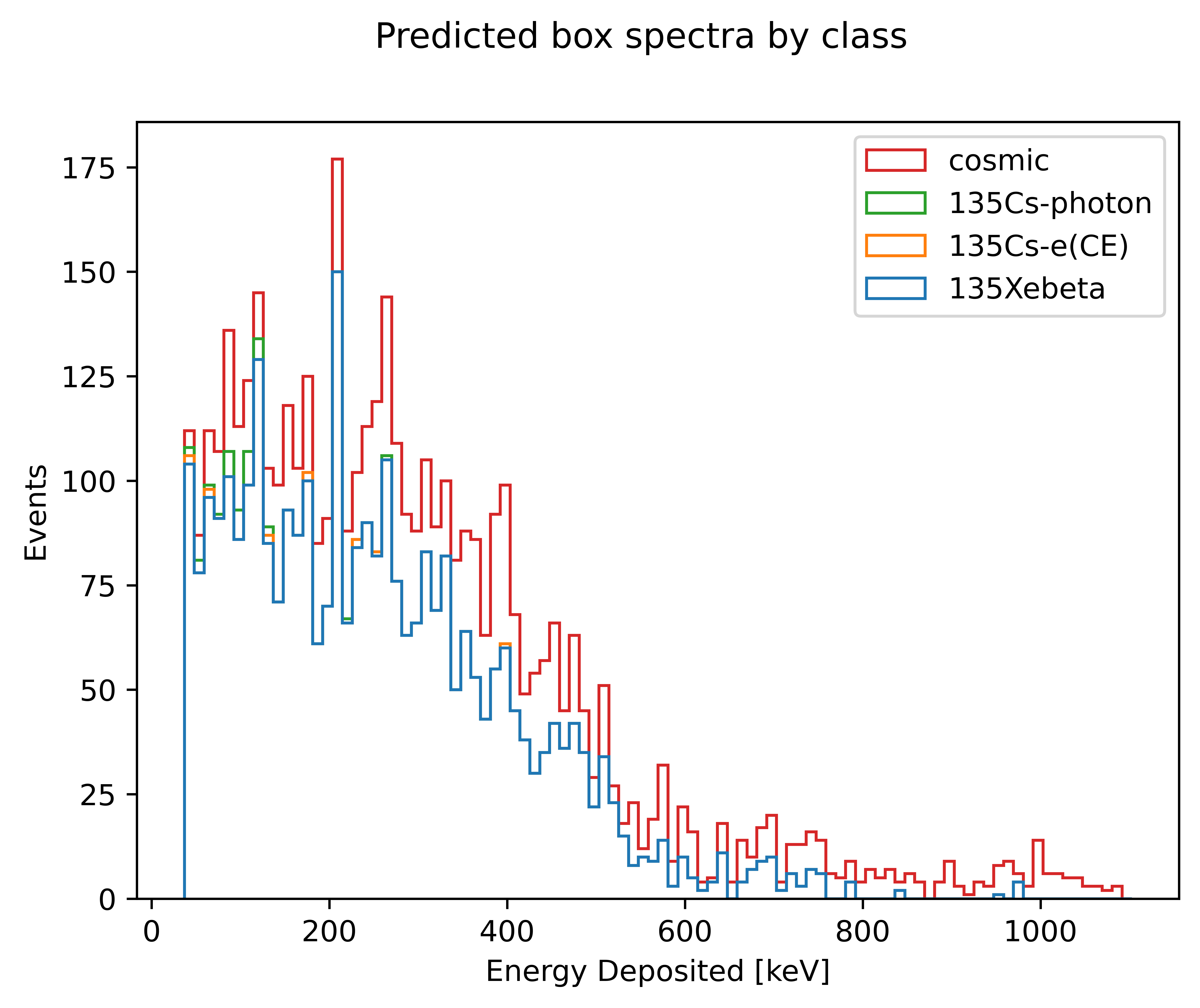}
 
        \caption{Shown is the stacked histogram of calculated energy distributions for the  network predictions, this time for a minimum score of 0.20. (The Truth distribution is the same as in Figure~\ref{fig:spectra}) The higher minimum score  has the effect of just allowing prediction of betas and cosmics (note nearly complete absence of green and orange).}
        \label{fig:spectra20}
    \end{figure}

Our predicted spectrum  shows two dominant issues we'd like to explore. We know that our CEs/gammas and betas are largely indistinguishable, and, second, we see that we're over-zealously producing cosmic muon bounding boxes. Thus, we wish to collapse all the electron-like activity, since the signatures are mainly topologically indistinct from one another, into one category and also only identify cosmics if they have higher predicted scores.  We find requiring minimum predicted scores of 0.20 achieves what we wish for: collapsing CEs/gammas/betas, at a cost, and also creating fewer extraneous muons. The cost is that, of course, we end up skipping over some activity of all types which otherwise would have been bound by boxes. Figure~\ref{fig:spectra20} shows the resulting spectra. We also observe that while we drop many clear redundant muon boxes, as desired, we are obviously now missing muons due to this new higher minimum score requirement. We also pursued this exercise more correctly, by properly altering the labels of the CEs/gammas and betas so that they are all one and the same, retrained the network, kept the minimum predicted score at 0.05;  much the same conclusions followed, though with higher efficiency/recall for the electromagnetic activity, as expected.

\subsection{Metrics for Object Identification}

We follow reference~\cite{PQmetric} for our various studies and direct the interested reader there to allow comparison to the community's more conventional object identification tasks. We discuss the panoptic quality (PQ) metric in this section.

A true positive (TP) box is defined as a  predicted box whose maximum IoU with a ground truth box exceeds some threshold. False negatives (FNs) are true boxes of activity which are never boxed in the prediction, according to the previous IoU requirement. This is missed activity. False positives are predicted boxes which are labeled by the wrong label, per our same IoU definition.

We define the panoptic quality metric in equation~\ref{eqn:PQ}. It is the product of correct-pixel-weighted  true positive boxes in an image (called segmentation quality) and the sum of true positives as a fraction of all true positives, false positives and false negatives (called recall quality). A given value of IoU defines correctness for the  predicted and groundtruth bounding boxes. We will investigate thresholds of {0.1,0.5,0.8} for IoU. One can identify $PQ$ in the particle/nuclear physics parlance as a product of purity and efficiency of some post-cuts sample.

    \begin{align}
        SQ &= \frac{\sum_{pixels-TP-bbox}IoU}{N_{TP}} \nonumber \\
        RQ &= \frac{N_{TP}}{N_{TP}+0.5*N_{FP} + 0.5*N_{FN}}\nonumber\\
         PQ &= SQ * RQ 
        \label{eqn:PQ}
    \end{align}

\begin{figure}
        \centering
        \includegraphics[width=4.0in]{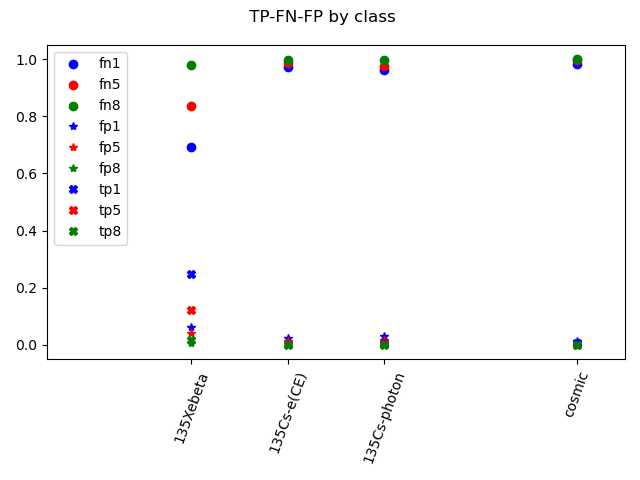}
        \includegraphics[width=4.0in]{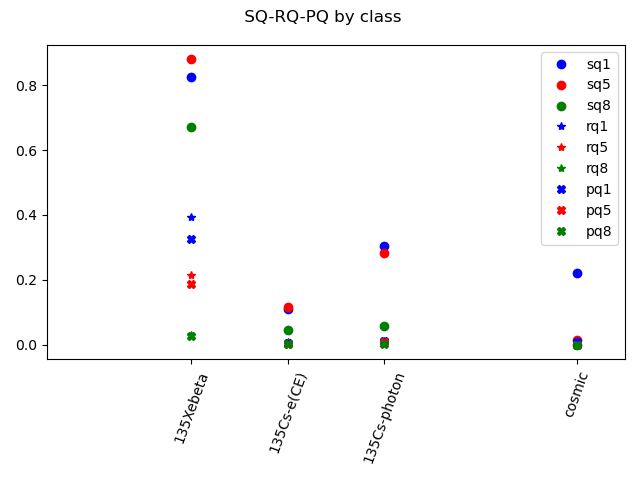}
        \caption{
        Shown are the TP,FP,FN metrics and SQ,RQ,PQ which follow from them. Please see text for meaning of each. The numerals 1,5,8 label the IoU requirement 0.1, 0.5, 0.8. Each metric is calculated separately for the four contributions shown. TP,FP,FN sum to unity separately for each contribution for a given IoU. Minimum score to appear in the prediction distribution is 0.05. The three colors show the metrics colored for different values of IoU. It is evident that IoU=0.10 is sufficient and offers best efficiency of getting true positives, without much negative impact from false positives or false negatives.}
        \label{fig:metrics}
    \end{figure}

\begin{figure}
        \centering
        \includegraphics[width=4.0in]{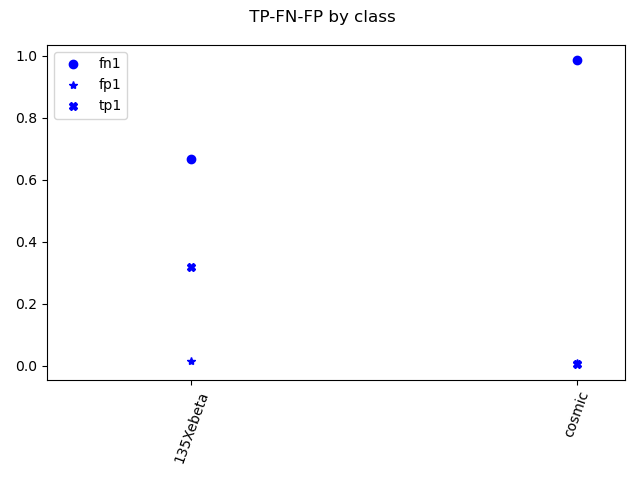}
        \includegraphics[width=4.0in]{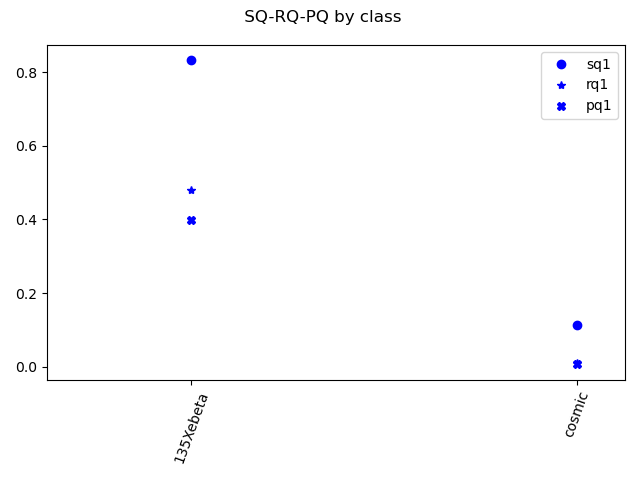}
        \caption{
        Shown are the TP,FP,FN metrics and SQ,RQ,PQ which follow from them. Please see text for meaning of each. In contrast to Fig~\ref{fig:metrics} we show only IoU=0.10 metrics. As before, TP,FP,FN sum to unity separately for each contribution. Minimum score to appear in prediction distribution is 0.20. Since this higher minimum score has the effect of collapsing all electromagnetic signatures into beta activity, we show only predicted betas and cosmics.}
        \label{fig:metrics20}
    \end{figure}

It's helpful to see the raw true positive, false positive, false negative scores that are ingredients to the panoptic quality and its components for our validation sample images. These ingredients are all shown in Figure~\ref{fig:metrics}  for the default minimum predicted class score of 0.05, and relevant metrics are shown again in Figure~\ref{fig:metrics20}  for the minimum prediction score of 0.20. The colors correspond to different IoU choices, which we remind, is what sets the definition for true/false predictions. Note first that the higher the IoU requirement in Figure~\ref{fig:metrics}, the  false negatives (missed activity of a particular type) go up quickly, while false positives only drop slowly. Therefore, we focus on the blue IoU=0.10 points.  This IoU choice causes us to miss the least actual activity while still capturing the correctness requirement of identified activity.

We note in comparing Figure~\ref{fig:metrics} and Figure~\ref{fig:metrics20}   that the higher required minimum prediction score has the effect of increasing the true positives for the betas while also decreasing the false positives. False negatives also go up for the betas, as a fraction of the total, but slowly. The net effect, comparing the corresponding PQ figures, shows we increase our PQ of 0.33 to 0.40 for betas with IoU=0.10. The cosmics continue to perform poorly under both minimum score requirements, as the root problem is still "boxing up" their activity completely and without redundancy.

\subsection{Diagnostic Sample: electrons}    
    In  this section we investigate if the relatively low numbers of conversion electrons and X-rays, for which PQ performance is not as high as that of the betas in the radioxenon sample,  might be classifiable on their own without the context of the full decay chain in the radioxenon. We also investigate if electrons launched from the front/back surfaces (mimicking actual radioactive decay electrons from the gas sample, imagined to be on either side of the CCD) are more or less classifiable than those in the bulk of the CCD (which mimic gammas that penetrate into the CCD and then undergo compton scattering).
    
    For this study we generate electrons in the range 100-200 keV. Further, in order to discover if the Bragg peak and diffusion effects aid or impair classification, we launch the electrons from the CCD front, back and bulk volume, in equal numbers. Electrons in this sample are launched with an isotropic initial direction into a $2\pi$ ($4\pi$, in the case of the bulk sample) solid angle. Electrons launched in the bulk can be considered as X-ray interactions.
    
\begin{figure}
        \centering
        \includegraphics[width=4.0in]{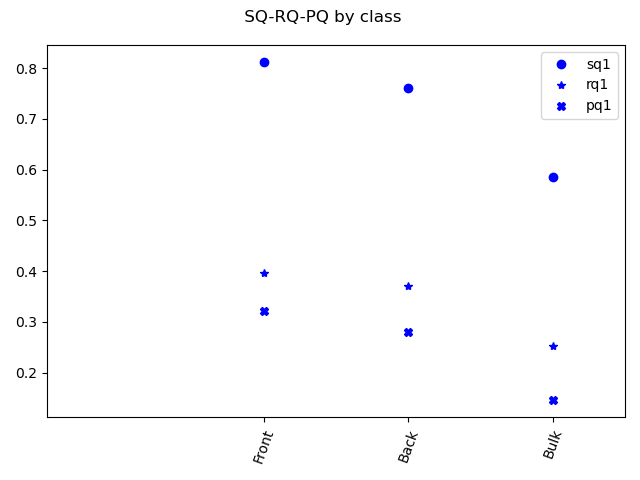}
        
        \caption{
        Shown are the SQ,RQ,PQ for our diagnostic 100-200 keV electron sample, with equal numbers launched from the back, front or bulk of the CCD. Minimum score to appear in prediction distribution is the standard 0.05.}
        \label{fig:metrics_es}
\end{figure}

\begin{figure}
        \centering
        \includegraphics[width=4.0in]{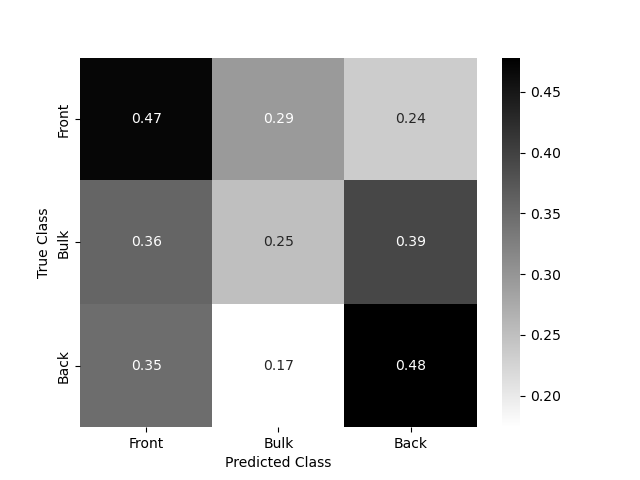}
        
        \caption{
        Confusion matrix for our diagnostic 100-200 keV electron sample, with equal numbers of electrons launched from the back, front and bulk of the CCD. Front and Back electrons are least confused by their competitors. IoU=0.1 for this matrix. }
        \label{fig:confusion_es}
\end{figure}

    We show in Figure~\ref{fig:metrics_es} that our efficiencies (rq1 in the plot) for properly classifying front and back electrons on the 100-200 keV (quite point like) sample are about 40\%, whereas we do worse for the bulk electrons, suggesting the network struggles to identify compton/pair-production from X-rays interacting deep in the silicon. Random guessing with no False Negatives would give 33\% for RQ. Meanwhile, once classified properly, our boxing-up is about 80 \% pure, leading to values for PQ $\approx$ 30\% (15\%) for front/back (bulk) electrons. We attribute the superiority of PQ for the front and back electrons to the contrast between the Bragg peak and "track" that is or is not diffused into an unhelpful blob, while the bulk electrons fall somewhere in between and lead to more random classification guessing. Figure~\ref{fig:confusion_es} shows the same story for the more commonly encountered confusion matrix. Correct identifications don't penalize for false negatives as in the PQ metric and its ingredients, so simple classification rates are higher. 
 
\section {Conclusion}
We demonstrate that thick CCDs exposed to radioactive noble gas isotopes can be analyzed with the Deep Learning technique of panoptic segmentation. 

We simulate a 1 Bq sample of \xe{135} and  cosmic activity on a 0.5mm thick CCD for a six minute exposure, and provide a realistic detector response. We attribute pixel-by-pixel ground truth as needed for the Detectron2 network, which to date has been mainly employed to identify macroscopic images in photographs. We show that while room for refinement exists, we can extract the most interesting result with reasonable efficiency and little distortion -- the energy spectrum of the betas.

We train a network with our simulated CCD exposures from a dedicated sample of such images. We apply a range of training, inference and analysis parameters, and we arrive at encouraging results in a test sample. While our main result is that the beta decay spectrum is discovered to be largely faithful to the expected one, we make other interesting observations. 1.) We find that requiring merely IoU=0.10 to define correctness of identification is suitable. 2.) Topologically indistinct species are unsurprisingly found to be predicted with some confusion among each other. 3.) Background muon tracks, while not our highest priority, are currently not well tagged and, we postulate, will be able to be better identified and subtracted with further optimization of network parameters. 4.) We find that  requiring  higher scores  for our predicted masks has the effect of collapsing all the electron-like species into one topologically identical category, and results in a slightly higher PQ score for this new blurred category, whereas cosmics struggle to meet this higher score and are not benefited.
    
A study of electrons launched from the front/back/bulk of the CCD leads us to see that there is some information from the track topology that the network capitalizes on. However, this exercise affirms our belief that most of the identification power for the radioxenon sample comes from the energy of the electron and perhaps regrettably, even its prior on the relative species' abundance in the training sample.

Generally, we find panoptic segmentation is a promising technique for CCD analysis for nuclear physics and dark matter search applications which, in the future, we wish to extend to other simulated radioisotope analyses and eventually to real CCD lab exposures to identify radioisotopes of interest.

\acknowledgments
This research was funded by the National Nuclear Security Administration, Defense Nuclear Nonproliferation Research and Development (NNSA DNN R\&D). The authors acknowledge important interdisciplinary collaboration with scientists and engineers from LANL, LLNL, PNNL, and SNL. PNNL is operated by Battelle Memorial Institute  for  the  U.S.  Department  of  Energy  (DOE) under Contract No. DE-AC05-76RL01830.

This research used resources of the Oak Ridge Leadership Computing Facility at the Oak Ridge National Laboratory, which is supported by the Office of Science of the U.S. Department of Energy under Contract No. DE-AC05-00OR22725.

\bibliographystyle{model1-num-names}
\bibliography{main_CCD-DL}

\end{document}